\documentclass[letter]{aa}  
\usepackage{graphicx,amsmath,wasysym}
\usepackage{amssymb}
\usepackage{color}
\usepackage{natbib}             
\usepackage{url}
\usepackage{multirow}
\bibpunct{(}{)}{;}{a}{}{,} 
\usepackage{sidecap}

\usepackage{array}
\usepackage{tabularx}
\usepackage{txfonts}
\newcolumntype{Z}{>{\centering\let\newline\\\arraybackslash\hspace{0pt}}X}
\newcommand{\lya}{Ly$\alpha$}
\newcommand{\Mearth}{\mathrm{M}_\oplus}
\newcommand{\Rearth}{\mathrm{R}_\oplus}
\newcommand{\hst}{\emph{HST}}
\newcommand{\kms}{km~s$^{-1}$}
\def\approxinf{%
  \def\p{%
    \setbox0=\vbox{\hbox{$<$}}%
    \ht0=0.6ex \box0 }%
  \def\s{%
    \vbox{\hbox{$\sim$}}%
  }%
  \mathrel{\raisebox{0.7ex}{%
      \mbox{$\underset{\s}{\p}$}%
    }}%
}

\def\approxsup{%
  \def\p{%
    \setbox0=\vbox{\hbox{$>$}}%
    \ht0=0.6ex \box0 }%
  \def\s{%
    \vbox{\hbox{$\sim$}}%
  }%
  \mathrel{\raisebox{0.7ex}{%
      \mbox{$\underset{\s}{\p}$}%
    }}%
}

\begin{document}

    \title{The long egress of GJ~436b's giant exosphere}
                               
   \author{
   B.~Lavie\inst{1,2},
   D.~Ehrenreich\inst{1},
   V.~Bourrier\inst{1},
   A.~Lecavelier des Etangs\inst{3},
   A.~Vidal-Madjar\inst{3},
   X.~Delfosse\inst{5},
   A.~Gracia~Berna\inst{2},
   K.~Heng\inst{2},
   N.~Thomas\inst{2},
   S.~Udry\inst{1}
   \&~P.~J.~Wheatley\inst{4}
   }
   
\authorrunning{B.~Lavie et al.}
\titlerunning{Egress of GJ~436b exosphere}


\institute{
Observatoire de l'Universit\'e de Gen\`eve, 51 chemin des Maillettes, 1290 Sauverny, Switzerland\and University of Bern, Space Research and Planetary Sciences, Sidlerstrasse 5, CH-3012, Bern, Switzerland\and CNRS, Institut d'Astrophysique de Paris, UMR 7095, 98bis boulevard Arago, 75014 Paris, France. \and Dept. of Physics, University of Warwick, Gibbet Hill Road, Coventry CV4 7AL, UK\and Univ. Grenoble Alpes, CNRS, IPAG, F-38000 Grenoble, France}
   
   \date{} 

  \abstract
{The M dwarf GJ\,436 hosts a transiting warm Neptune known to experience atmospheric escape. Previous observations revealed the presence of a giant hydrogen exosphere transiting the star for more than 5~h, and absorbing up to 56\% of the flux in the blue wing of the stellar Lyman-$\alpha$ line of neutral hydrogen (\ion{H}{i} \lya). The unexpected size of this comet-like exosphere prevented observing the full transit of its tail. In this Letter, we present new \lya\ observations of GJ~436 obtained with the \emph{Space Telescope Imaging Spectrograph} (STIS) instrument onboard the \emph{Hubble Space Telescope}. The stability of the \lya\ line over six years allowed us to combine these new observations with archival data sets, substantially expanding the coverage of the exospheric transit. Hydrogen atoms in the tail of the exospheric cloud keep occulting the star for 10--25~h after the transit of the planet, remarkably confirming a previous prediction based on 3D numerical simulations with the \texttt{EVaporating Exoplanet} code (EVE). This result strengthens the interpretation that the exosphere of GJ~436b is shaped by both radiative braking and charge exchanges with the stellar wind. We further report flux decreases of $15\pm2\%$ and $47\pm 10\%$ in the red wing of the \lya\ line and in the line of ionised silicon (\ion{Si}{iii}). Despite some temporal variability possibly linked with stellar activity, these two signals occur during the exospheric transit and could be of planetary origin. Follow-up observations will be required to assess the possibility that the redshifted \lya\ and \ion{Si}{iii} absorption signatures arise from interactions between the exospheric flow and the magnetic field of the star.}


   \maketitle

\section{Introduction}
\label{intro} 
Transit observations in the stellar Lyman-$\alpha$ line of neutral hydrogen (\ion{H}{i} \lya\ at 1\,215.67~\AA) allowed the detection of atmospheric escape from the two hot Jupiters HD~209458b \citep{Vidal-Madjar2003,Vidal-Madjar2004} and HD~189733b \citep{Lecavelier2010,Lecavelier2012,Bourrier2013b}, the warm giant 55~Cnc~b \citep{Ehrenreich2012}, and the warm Neptune GJ~436b \citep{Kulow2014,Ehrenreich2015a}. The escaping hydrogen exospheres produce large \lya\ absorption transit signatures ranging from $7.5\pm1.8\%$ for 55~Cnc~b to $56.3\pm3.5\%$ for GJ~436b. Compared to close-in hot Jupiters, GJ 436b ($25.4\pm2.1~\Mearth$ and $4.10\pm0.16~\Rearth$, \citealt{Butler2004,Gillon2007,Lanotte2014a}) is gently irradiated by its M2{\sc v} host star \citep{Ehrenreich2011b}, yet it spots an extended hydrogen envelope, first hinted at by \citet{Kulow2014} and fully revealed by \cite{Ehrenreich2015a}. These authors found that GJ~436b emits a comet-like cloud of H atoms, with a coma bigger than the star and a  tail extending millions of kilometers (up to 40\% of GJ~436b's orbit,~ 450 planetary radii), trailing the planet. Because of its unexpected, gigantic scale, past \lya\ observations of GJ~436b's exosphere could only cover the transit of its coma and the onset of its comet-like tail. \citet{Bourrier2015a,Bourrier2016a} have modelled the dynamics of GJ~436b exospheric cloud using the \texttt{EVaporating Exoplanets} code \citep[\texttt{EVE;}][]{Bourrier2013}. Adjusting \lya\ spectra obtained between 2010 and 2014, they show that the geometry and dynamical structure of the exospheric cloud could be explained by radiative braking (i.e. the effect on exosphere particles resulting from radiation pressure lower than stellar gravity; \citealt{Bourrier2015a}) and stellar wind interaction (specifically, charge exchange; \citealt{Bourrier2016a}). The balance between both mechanisms, however, would shape the cloud tail, hence the egress of the UV transit, differently. Meanwhile, the full extent of the tail, provided by the egress duration, is unknown. Given the partial coverage (3--4h) of the egress, 
new data were needed to fully cover the UV transit egress and precisely determine the nature of the star-planet interaction sculpting the cloud.

In this Letter, we present new \emph{Hubble Space Telescope} (\hst) data obtained with the Space Telescope Imaging Spectrograph (STIS). The combined reduction and analysis of these data covering a much larger fraction of the planetary orbit, with all previously existing data, is presented in Sect.~\ref{sec:analysis} and Appendix~\ref{appendix_sys_corr}. Our results (Sect.~\ref{analysis_exosphere}) confirm the blueshifted \lya\ absorption signature of \citet{Ehrenreich2015a} and strengthen the interpretation of \citet{Bourrier2015a,Bourrier2016a}. We also find new surprising absorption signatures in the red wing of the \ion{H}{i} \lya\ line and in the ionised silicon line (\ion{Si}{iii}) that could be of planetary or stellar origins.


\section{Observations} 
\label{sec:obs}

In total, GJ~436b has been observed at eight epochs (hereafter Visits 0 to 7) with \hst/STIS. All visits are listed in Table~\ref{tab:data}. Visits 0 to 3 revealed the deep transit of GJ~436b's exosphere in the blue wing of the \lya\ line \citep{Ehrenreich2011b,Ehrenreich2015a,Kulow2014}. Three new visits (Visits 5 to 7) were obtained on 2016-Mar-30, 2016-Apr-06, and 2016-May-08 (General Observer (GO) programme 14222; PI: D.~Ehrenreich) with the aim of completing the coverage of the exospheric transit, during the transit of the tail ($\sim$+3,+20~h; Visit 6), before the transit of the coma ($\approxinf -3$~h; Visit 7), and after the presumed end of the transit ($\approxsup$+20~h; Visit 5). Times are calculated using the ephemeris from \citet{Lanotte2014a}. All observations were made with the Far Ultraviolet Multi-Anode Microchannel Array detector (FUV-MAMA) detector and the G140M grating with a central wavelength of 1\,222~\AA. 
The eight visits represent 27 \hst\ orbits. Each orbit-long\footnote{The exposure time ranges from $\sim1\,500$ to $2\,900$~s.} time-tag exposure has been divided into five sub-exposures (a reasonable balance between signal over noise ratio and time resolution)  and processed through \texttt{CALSTIS}, the STIS pipeline, yielding a total of 135 spectra.

\section{Analysis}
\label{sec:analysis}

\begin{figure}
\centering
\includegraphics[width=\columnwidth,trim={0 1.3cm 0 0}]{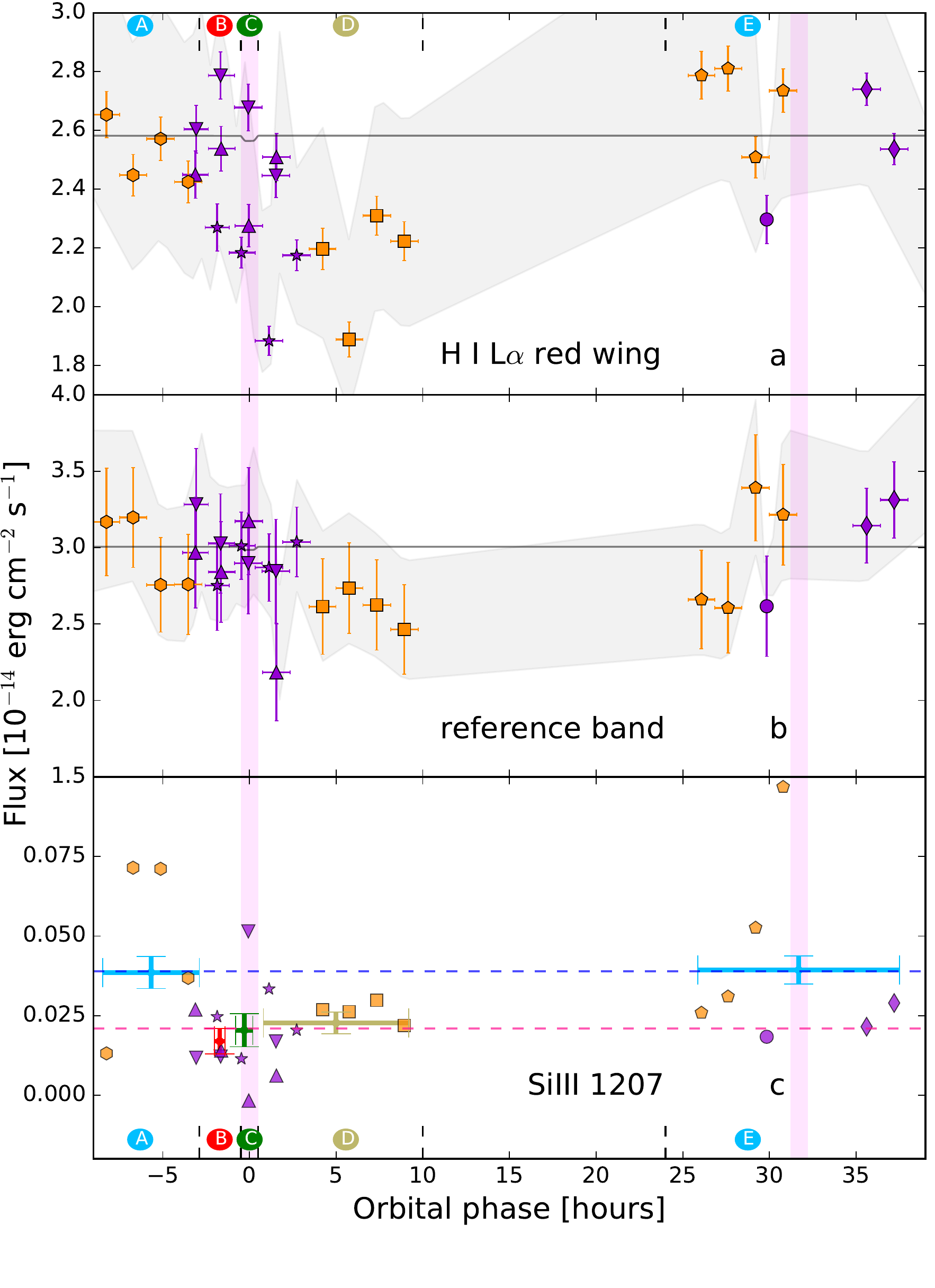}
\caption{\label{lightcurve} Light curves of GJ~436 integrated \lya\ red wing (a), \lya\ reference band $[-250,-120]\cup[+120,+250]$~\kms\ (b) and \ion{Si}{iii} line (c). 
New visits described in this work are in orange while previous visits are plotted in violet. Symbols are in Table~\ref{tab:data}. Different temporal regions are defined: (A) before transit, (B) ingress, (C) optical transit, (D) egress, and (E) after transit. The grey-filled region represents the $1\sigma$ confidence interval of the systematic correction method using the Gaussian processes (Appendix.~\ref{appendix_sys_corr}). The vertical magenta zones show the optical primary and secondary transit. The optical transit light curve of GJ~436b is indicated with the black line. In panel (c), fluxes are integrated for the different temporal region (see Fig~\ref{lc_blueLya}). Horizontal dashed lines indicate the out-of-transit flux (blue - regions A and E) and the in-transit flux (pink - regions B, C, and D)}
\end{figure}

\subsection{Reference, unocculted \lya\ flux}
The new visits obtained before and after the exospheric transit allow us to reconstruct a reference spectrum seemingly unaffected by the exospheric absorption signature. We average the flux in the first \hst\ orbits of Visits 0, 2, and 3 \citep[as done by][]{Ehrenreich2015a} with the flux measured in all \hst\ orbits obtained during Visits 4, 5, and 7. This average out-of-transit baseline spectrum appears stable in both the red wing of the \lya\ line and the reference band ($[-250,-120]\cup[+120,+250]$~\kms), as can be seen in Fig.~\ref{lightcurve}a,b. 
In the \lya\ blue wing (Fig.~\ref{lc_blueLya}), the flux of Visit~4 and the first exposure of Visit~5 are higher than in subsequent orbits used for the out-of-transit baseline. The origin of this wavelength-dependent rise is unclear, but in the following we chose to include them in the out-of-transit baseline flux. Excluding them will only decrease the out-of-transit baseline by less than 4\%.

\subsection{Correction of systematics}
In addition to stellar variability, STIS G140M spectra are known to be impacted by an instrumental systematic effect caused by the telescope breathing. 
This effect is reported to be achromatic, thus to correct for it we need to locate a reference wavelength or velocity band in the \lya\ emission feature devoid of astrophysical signal. The geocoronal emission line (airglow) contaminates the core of the observed stellar line, and varies in strength and position from one observation to another. This region is not adapted to our needs so we exclude it, setting conservative limits of $[-40,+30]$~\kms\ (velocities are expressed with respect to the \lya\ line centre in the stellar rest frame). Previous work reported absorption signatures in the \lya\ blue wing \citep[${[}-120,-40{]}$~\kms;][]{Ehrenreich2015a} and, tentatively, in the \lya\ red wing \citep[${[}+30,+120{]}$~\kms;][]{Kulow2014}. These two bands, over which we will search for exospheric signatures, cannot be used to monitor telescope breathing. A careful inspection of the \lya\ spectra presented in Fig.~\ref{spectrum_timelapse} allowed us to find two bands seemingly free of astrophysical signal, over $[-250,-120]$ and $[+120,+250]$~\kms, which we call the reference band. The flux integrated over these two bands, which can be seen in Fig.~\ref{lightcurve_raw}c, remains about stable within the uncertainties for all visits, spanning over six years of observations. We hypothesized that all variations in these reference bands could be attributed to instrumental effect. Different strategies have been developed to correct for telescope breathing \citep[see e.g.][and references therein]{Bourrier2017}. In this work, we compare three correction methods: parametric \citep{Bourrier2013,Ehrenreich2012,Bourrier2016b}, empirical \citep{Deming2013,Wilkins2014,Berta2012,Ehrenreich2014}, and non-parametric using Gaussian processes (GP; Appendix~\ref{appendix_sys_corr}). All three methods yield similar results. In the following, we opt to rely on the GP-based approach.

\subsection{Flux in other spectral lines}
\label{other_flux_analysis}
Besides the prominent \lya\ line, we detect the following stellar emission features in the STIS/G140M range (1\,190--1\,250~\AA):  \ion{Si}{iii} at 1\,206.5~\AA, \ion{N}{v} doublet at 1\,238.8 and 1\,242.8~\AA, and \ion{O}{V} at 1\,218.3~\AA\ (cf.~Table~\ref{tab:otherlines}). The low signal-to-noise ratio in these lines prevented us from doing an orbit-to-orbit comparison and establishing the stability of the stellar baseline flux for those lines. We therefore averaged  their spectra within the orbital phase regions labelled A to E on Fig.~\ref{lc_blueLya} (cf.~Table~\ref{tab:defregion}): (A) before the exospheric transit signature in the blue wing of \lya, (B) during the exospheric transit ingress, (C) during the optical transit (i.e. the transit of the planetary disk alone), (D) during the exospheric transit egress, and (E) after the exospheric transit. Regions (B), (C), and (D) are considered `in transit'; regions (A) and (E) are `out of transit'.


\section{Results}
\label{analysis_exosphere}

\subsection{Detection of the \lya\ transit egress} 
\label{bluewing}
\begin{figure}  
\includegraphics[width=\columnwidth]{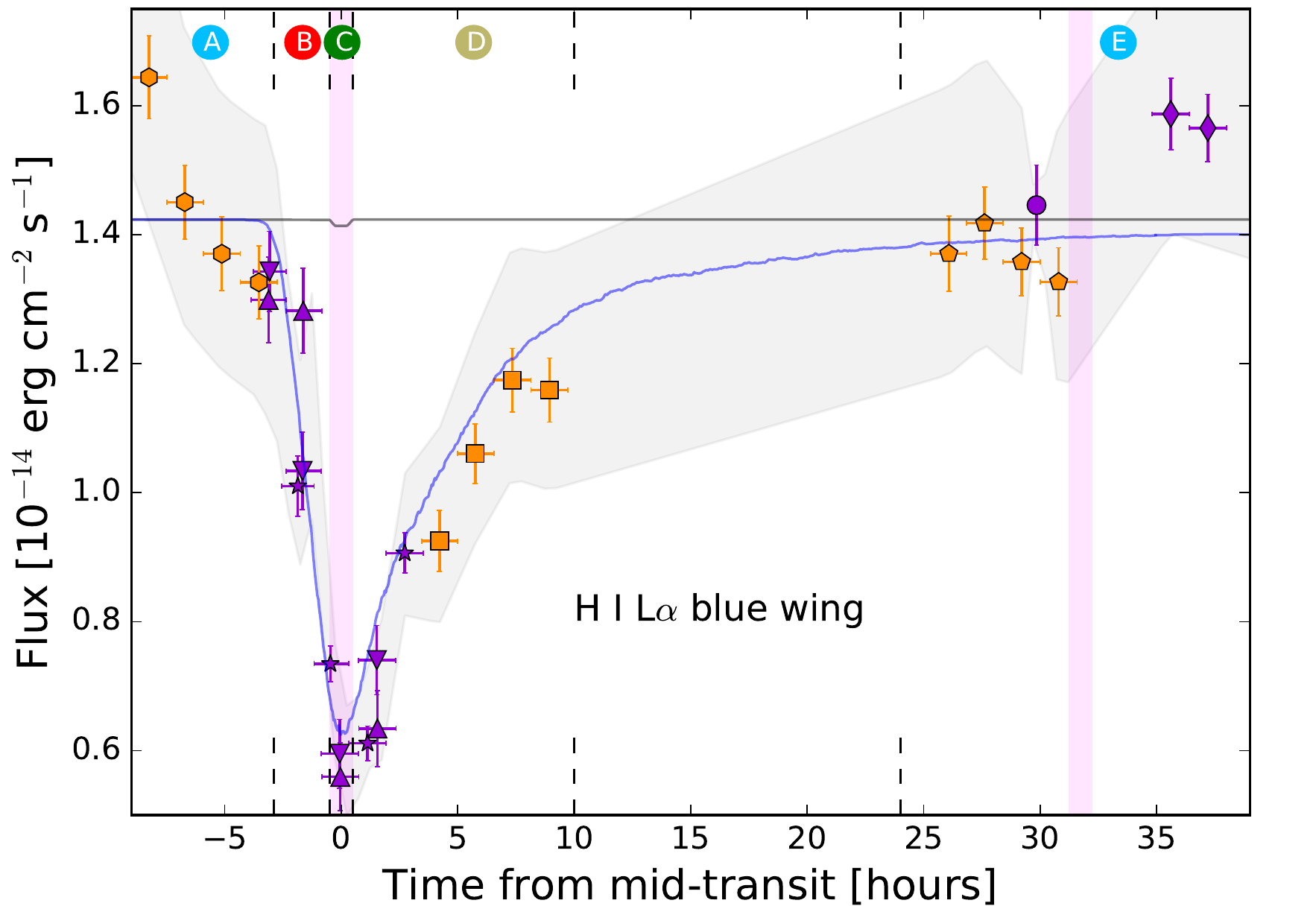}
\caption[]{\label{lc_blueLya} Light curves of GJ~436 integrated over the \lya\ blue wing. Legend is the same as in Fig~\ref{lightcurve}. The blue curve is the model calculated with \texttt{EVE} that represents the best fit to previous data (Visit 0 to 3).
}
\end{figure}

 The most significant absorption signal is detected in the \lya\ blue wing between $-120$ and $-40$~\kms\ (Fig.~\ref{lc_blueLya}), in agreement with \citet{Ehrenreich2015a}. Visit~6 (time from mid-transit $\tau = [+4.01,+9.18]$~h) prolongates and confirms the exospheric egress suggested by Visits~1, 2, and~3. The egress duration is now constrained to be longer than $\sim$10~h and shorter than $\sim$25~h. The \lya\ flux measured during Visit~5 ($\tau = [+25.8,+31.0]$~h) is in good agreement with the Visit~0 measurement obtained $\sim$+30~h after the mid-transit time, close to superior conjunction. We stress that those compatible measurements were obtained more than six years apart, highlighting the temporal stability of GJ~436.
\citet{Bourrier2015a,Bourrier2016b} fitted the spectra obtained during Visits~0 to 3 (all violet curves in Fig.~\ref{spect_timeline}) with \texttt{EVE}. Their best-fit theoretical spectra yield the light curve plotted in Fig.~\ref{lc_blueLya}. The model is in remarkable agreement with the new data from Visits~6 and~7, strengthening the interpretation proposed in \citet{Bourrier2015a,Bourrier2016b}: the weak UV radiation emitted by GJ~436 yields a low radiation pressure ($\sim$70\% of stellar gravity) and a low photoionisation of escaping hydrogen atoms. This, combined with a high planetary wind velocity ($\sim$55~\kms), allows the planetary outflow to diffuse within a large coma surrounding and comoving with the planet, which further extends into a broad cometary tail \citep{Bourrier2015a}. 
In addition, escaping atoms interact with the slow stellar wind of the M dwarf ($\sim$85~\kms) via charge exchanges abrading the day-side exosphere \citep{Bourrier2016b}. These interactions create a secondary tail of neutralised stellar wind protons\footnote{These are stellar wind protons that gained an electron from the interaction with neutral hydrogen atoms in the exosphere.}, which move with the persistent dynamics of the stellar wind and therefore yield different \lya\ absorption signatures over time than the primary tail. 
This model assumed that the out-of-transit baseline is provided by the flux measured during the 'out-of-transit' region defined in the previous section. Visits~4 and 7 tentatively suggest (Fig.~\ref{lc_blueLya}), however, that the unocculted \lya\ line might be even brighter. This will have to be confirmed with additional measurements obtained between the superior conjunction and the first quadrature.

\subsection{Detection of a redshifted \lya\ absorption signature}
\label{redwing} 
\begin{figure}  
\includegraphics[width=\columnwidth, trim={0 0 0 0}]{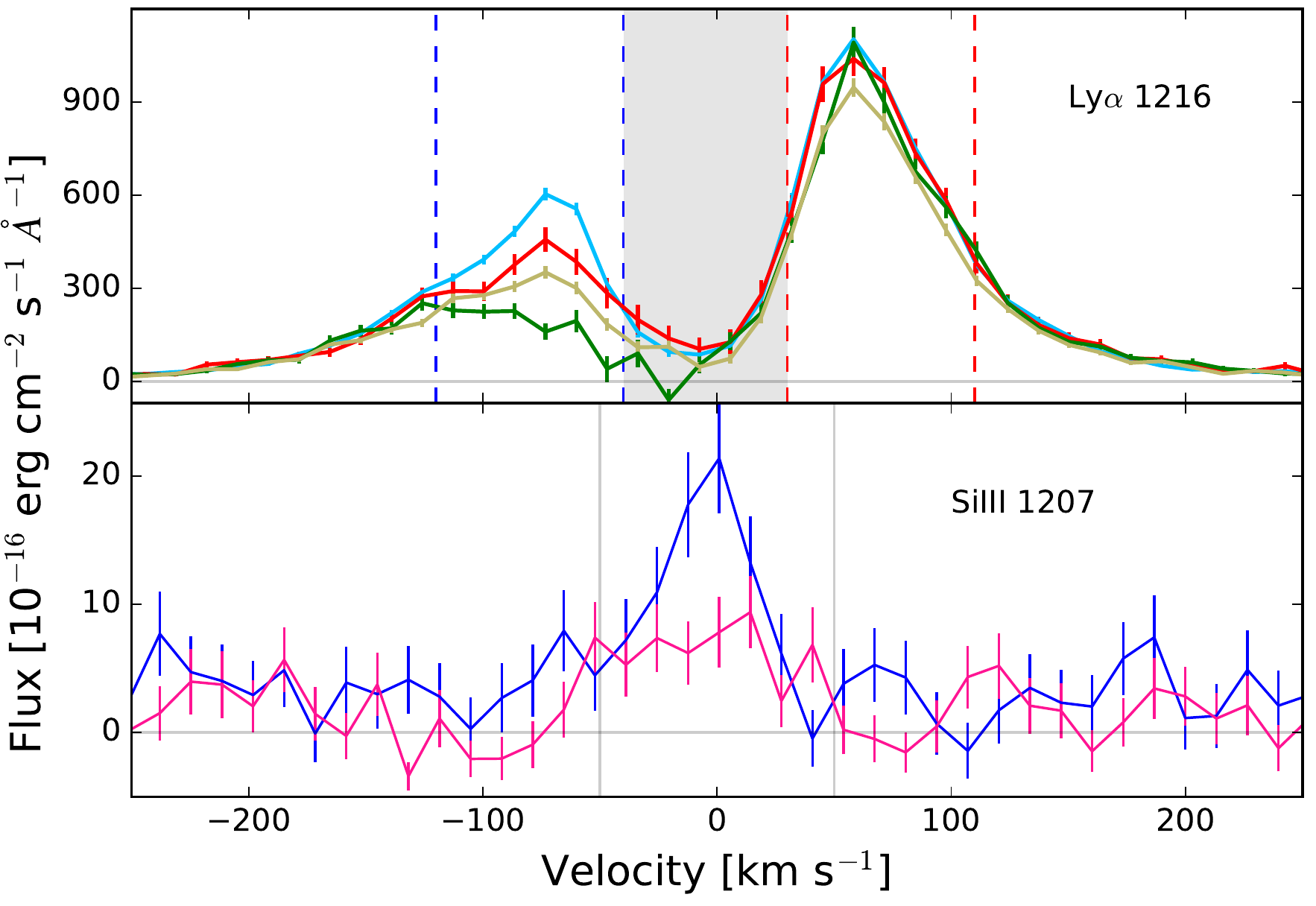}
\caption[]{\label{spectrum_timelapse} Top panel: Averaged out-of-transit (blue), ingress(red), egress (khaki), and in-transit (green) spectra of the \lya\ line at $1\,206.5$~\AA. The grey zone is the geocoronal emission (airglow) band.
Bottom panel : Averaged out-of-transit (blue) and in-transit (pink) spectra of the \ion{Si}{iii} at $1\,206.5$~\AA. Vertical grey line indicates the $[-50,+50]$~\kms area.}
\end{figure}

Visits~1 and~6 show similar integrated fluxes in the \lya\ red wing during and after the optical transit, with an average absorption depth of $15\pm2\%$ compared to the baseline level (Figs.~\ref{lightcurve}a \&~\ref{spectrum_timelapse}). In both visits, the absorption signature is located within $[+30,+110]$~\kms\ but at a different time from the mid-transit at +0.5 and +6.6~h, respectively. In comparison, the blueshifted absorption occurs around the mid-transit. This redshifted absorption is time variable. This was first noted by \citet{Kulow2014} from Visit~1 data; however, \citet{Ehrenreich2015a} did not confirm it with same-phase data obtained during Visits~2 and 3. Visit~6 changes this picture and confirms the existence of a redshifted \lya\ signature delayed in time with respect to the blueshifted signature. Interpretation of this redshifted \lya\ signature is challenging: the \texttt{EVE} model that best fitted the blueshifted signature (Sect.~\ref{bluewing}) does predict the existence of H atoms moving towards the star, but redshifted by less than +50~\kms\  because this population is localised within the coma and could not create a signature in the tail \citep[see Fig.~4 in][]{Bourrier2016b}. 
In the absence of radiation pressure or with a very strong self-shielding protecting the interior of the coma (which is not the case for GJ436b), \citet{Bourrier2015a} showed that gravitational shear could lead to a stream of H atoms falling onto the star  but ahead of the planet and thus yielding a redshifted signature before the optical transit (see their Fig 2).
Interestingly, tentative \lya\ redshifted absorptions have been previously reported in HD~209458b and HD~189733b \citep{Vidal-Madjar2003,Lecavelier2012} but in phase with the respective blueshifted absorptions signature of those planets.\\
It is plausible that stellar variability \citep{Tian2009,Llama2016,Vidal1975} is the origin of the variations, because they are located at the peak of the observed red wing, which traces higher altitudes in the chromosphere of the star which are more active than the lower chromosphere which is the source for the far wings of the \lya\ line \citep[e.g.][]{Bourrier2017}. However, the high stability of the stellar flux over six years of observations, the fact that the variations always correspond to a decrease of the flux with respect to the reference spectrum, and the fact that they appear to be phased during and after the planet transit, suggest another scenario.\\
Interaction of the exospheric flow with the star and planet magnetic fields could be one of the mechanisms that drives particles towards the star. A bow shock formed ahead of the planet by the interaction between the stellar wind and the planetary magnetosphere could only yield absorption before the planet transit \citep{Alexander2016a,Lai2010a}. Furthermore, \citet{Bourrier2016b} showed that the interactions between the exosphere and the stellar wind required to explain the absorption in the \lya\ blue wing imply that the planetary magnetosphere is embedded deep inside the coma. 
Strong magnetic connection could nonetheless link the embedded magnetosphere with the stellar magnetic field, forcing part of the planetary outflow to stream toward the star behind the planet \citep{Strugarek2016a}. 
One of the magnetised star-planet interactions proposed by \citet{Matsakos2015a} (Type 4) bears some resemblance with our observations in the red wing, with planetary gas infalling nearly radially onto the star. The variability of the detected absorption could be explained by the continuous readjustment of the magnetic field topology as the planet orbits the star, with new magnetic accretion channels forming periodically. However, this scenario requires a high UV irradiation and a weak planetary outflow, which are not consistent with the mild irradiation from GJ~436 and the high velocities derived for the planetary outflow \citep{Bourrier2016b}. Furthermore, \citet{Matsakos2015a} investigated the case of a hot Jupiter around a solar-type star, without accounting for radiation pressure despite its major role in shaping the exospheres of evaporating exoplanets. Gas in magnetised interaction regions is also expected to be strongly ionised, which would require either that massive amounts of neutral hydrogen infall toward the star or that shielding or recombination mechanisms allow for a significant portion of this gas to remain neutral.

\subsection{Detection of an absorption signature from ionised silicon} 
\label{silicon_result}

The other stellar lines (Table~\ref{tab:otherlines}) are much fainter than \lya, so we compared for each line the total fluxes averaged in transit (phase regions B, C and D) and out of transit (phase regions A and E). We report an absorption signal of $36\pm 15\%$ in the \ion{O}{v} line. The signal occurs during ingress and egress but not during the optical transit. Thus, it may not be of planetary origin.
\citet{Parke-Loyd2017} observed GJ~436 with \hst/Cosmic Origins Spectrograph (COS) from $[-4,+3]$~h around the optical transit. Within this time range, they reported no detection of \ion{N}{v} or \ion{Si}{iii} absorption. This is in agreement with our analysis of the two \ion{N}{v} lines. Meanwhile, we detect an absorption signal of $47\pm10\%$ in the \ion{Si}{iii} line within $[-50,+50]$~\kms, similar in intensity to the blueshifted \lya\ absorption depth and occuring during the same phases as the exospheric transit. The flux is  constant over the phase range covered by \citet{Parke-Loyd2017} (Fig.~\ref{lightcurve} c), suggesting that the stellar line was already absorbed during their observations, preventing them from detecting any variations. We note that our observations are compatible with their upper limit.
We cannot exclude stellar variations in the \ion{Si}{iii} line \citep{Parke-Loyd2017}, which could be linked to the variation seen in the \lya\ red wing (Sect.~\ref{redwing}). A planetary origin for the \ion{Si}{iii} signal is tantalising, as it would demonstrate the hydrodynamic nature of GJ~436b atmospheric escape, constrain the star-planet magnetic interaction, and provide a possible tracer for the presence of enstatite clouds (Mg$_2$Si$_2$O$_6$), potentially responsible for the flat near-infrared transmission spectrum of the lower atmosphere \citep{Knutson:2014aa}. This will require additional observations.

\section{Conclusion}
We report new \hst/STIS UV observations of the exospheric cloud escaping from the warm Neptune GJ~436b. A combined analysis of all available UV data, making use of GP to correct for systematics, yields the following results: 
\begin{enumerate}
\item We detect the UV transit egress and constrain its duration between 10 to 25~h. This corresponds to a size of the exospheric hydrogen tail between 5 and 12 millions~km. This result confirms previous observations \citep{Ehrenreich2015a} and their interpretation \citep{Bourrier2015a,Bourrier2016b}.
\item We detect an absorption signal in the red wing of the \lya\ line, which is delayed in time compared to the blueshifted absorption. This signal could originate either from the planet or be due to stellar activity. 
\item We detect an absorption signal in the \ion{Si}{iii} line, possibly linked with the \lya\ redshifted signal (and stellar activity). More observations will be needed at other phases to discriminate the stellar activity scenario from a planetary origin.
\item We notice the remarkable stability of GJ~436's unocculted \lya\ emission line over the six-year period (2010--2016) covered by the available observations. 
\end{enumerate}

\begin{acknowledgements}
We thank the referee, Allison Youngblood, for useful and fruitful comments. This work is based on observations made with the NASA/ESA Hubble Space Telescope. This work has been carried out in the frame of the National Centre for Competence in Research `PlanetS' supported by the Swiss National Science Foundation (SNSF). B.L., D.E., V.B., K.H., N.T., and S.U. acknowledge the financial support of the SNSF. A.L. acknowledges financial support from the Centre National d\textsc{\char13}Etudes Spatiales (CNES). PW is supported by the UK Science and Technology Facilities Council under consolidated grant ST/P000495/1. The authors acknowledge the support of the French Agence Nationale de la Recherche (ANR), under programme ANR-12-BS05-0012 `Exo-Atmos'. This project has received funding from the European Research Council (ERC) under the European Union's Horizon 2020 research and innovation programme (project {\sc Four Aces}; grant agreement No 724427). 
\end{acknowledgements}

\bibliographystyle{aa} 
\bibliography{UV,Outils_stats}{} 

\begin{appendix}
\section{Log of observations and raw light curves}
\label{app:obs}
All spectra were linearly interpolated on a common wavelength grid, chosen to be the grid of Visit 0 first sub-exposure.

\begin{figure}
\centering
\includegraphics[width=\columnwidth,trim={0 2cm 0 0}]{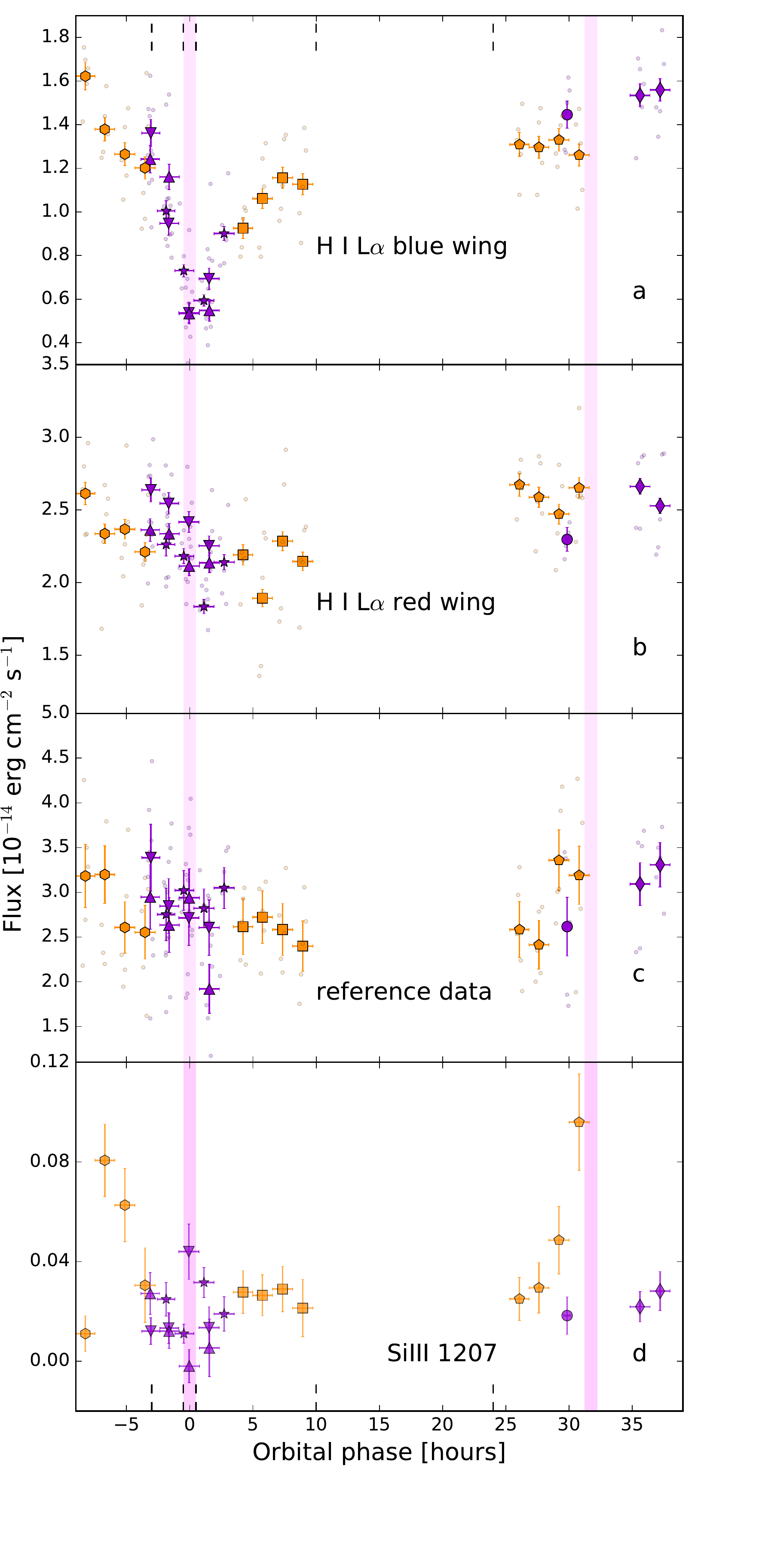}
\caption{\label{lightcurve_raw}\lya\ light curve of GJ~436 obtained from integrating the raw spectra uncorrected for systematics. The flux is integrated over the blue wing $[-120,-40]$~\kms\ (a), the red wing $[+30,+120]$~\kms\ ( b), the reference band $[-250,-120]\cup[+120,+250]$~\kms\ (c), and the  $[-50,50]$~\kms\ band of the \ion{Si}{iii} line (d). The colour code is the same as in Fig.~\ref{lightcurve}. Large coloured symbols with errors bars are the individual \hst\ orbits (see Table~\ref{tab:data} for the visit symbols) while small grey circles are the time-tagged sub-exposures extracted from each \hst\ orbit for panel a, b, and c.}
\end{figure}
\begin{figure}  
\centering
\includegraphics[width=\columnwidth,trim={0 3cm 0 0}]{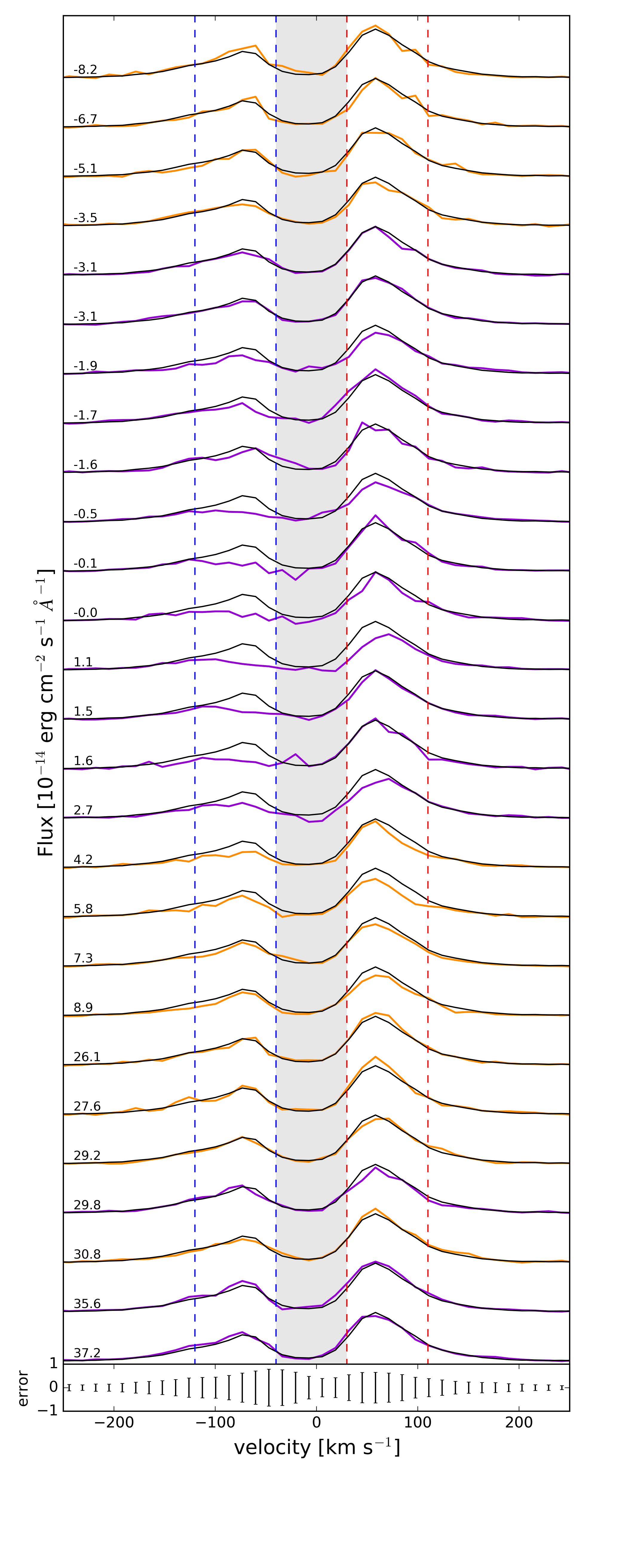}
\caption[]{\label{spect_timeline} New visits described in this work are in orange while previous visits are plotted in violet. \lya\ line spectra of GJ~436 of each HST orbit compared to the unocculted line (black curves) obtained by combining all spectra from the out-of-transit phases region (see Sect. \ref{other_flux_analysis}).The number on top of each spectrum indicates the time in hours from the mid-transit time. Vertical dashed lines indicate the blue and the red bands showing absorption signatures. The grey zone is the geocoronal emission (airglow) band.}
\end{figure}

\clearpage

\begin{table*}
\begin{center}
\caption{\label{tab:data}Observation log. Symbols in the first column refer to plotting symbols in Fig.~\ref{lightcurve_raw}.}
\begin{tabular}{ccllccc}
\hline
\hline
Visit                           & Date          & Programme               & PI                      & \hst\ orbits  & Time from mid-transit (h)     & Phase range             \\
\hline
0 ($\circ$)                     & 01 May 2010   & GO\#11817             & Ehrenreich      & 1                     & $[29.64,30.03]        $                                       & $[0.46,0.47]$                   \\
1 ($\star$)                     & 07 Dec 2012   & GTO\#12034            & Green           & 4                     & $[-2.03,3.04]$                                        & $[-0.03,0.05]$                  \\
2 ($\APLup$)            & 18 Jun 2013   & GO\#12965             & Ehrenreich    & 4                       & $[-3.26,1.75]$                                        & $[-0.05,0.03]$                  \\
3 ($\APLdown$)  & 23 Jun 2014   & GO\#12965             & Ehrenreich    & 4                       & $[-3.30,1.78]$                                        & $[-0.05,0.03]$                  \\
4 ($\diamond$)        & 25 Jun 2015     & GO\#13650             & France                & 2                       & $[35.29,37.49]$                                       & $[-0.44,-0.41]$                 \\
5 ($\pentagon$) & 30 Mar 2016   & GO\#14222             & Ehrenreich    & 4                       & $[25.86,31.04]$                                       & $[ 0.40,0.49]$                  \\
6 ($\square$)           & 06 Apr 2016   & GO\#14222             & Ehrenreich    & 4                       & $[4.01,9.18]$                                 & $[0.06,0.14]$                   \\
7 ($\varhexagon$)       & 08 May 2016   & GO\#14222             & Ehrenreich    & 4                       & $[-8.45,-3.28]$                                       & $[-0.13,-0.05]$                 \\
\hline
\end{tabular}
\end{center}
\end{table*}

\begin{table}
\begin{center}
\caption{\label{tab:defregion} Orbital phases range}
\begin{tabular}{ll}
\hline
\hline
Phase            & \hst\ orbits \\
range             & [visit \#, (orbit \#)]  \\
\hline
Before transit & [2,(1)], [3,(1)], [7,(1,2,3,4)]\\
Ingress & [1,(1)] , [2,(2)], [3,(2)]\\
Optical transit & [1,(2)] , [2,(3)], [3,(3)]\\
Egress & [1,(3,4)] , [2,(4)], [3,(3)],[6,(1,2,3,4)]\\
After transit & [0,(1)],[4,(1,2)],[5,(1,2,3,4)]\\
\hline
\end{tabular}
\end{center}
\end{table}

\clearpage
\section{Correction of systematics: Gaussian processes}
\label{appendix_sys_corr}
This method differs from the parametric approach in that the shape of the function used to describe the systematics can be adjusted to capture any behaviour not encompassed by an analytical function. Gaussian processes (GP) are one of these methods. GP are widely use in machine learning and are increasingly popular in the exoplanet community (see \cite{RasmussenWilliams2006} for a global introduction and \cite{Gibson2012} for an application in the exoplanet field). Within a GP scheme, the joint probability distribution for the reference band is a multivariate gaussian distributed about a mean function, a flat line representing a stable flux in our case. Systematics and white noise are characterised by the covariance matrix, which is defined by a covariance function (or kernel). 
In this analysis we adopt the squared exponential kernel in addition to a white kernel: $k(t,t^{'}) =  \exp(\frac{-{\lVert}{t'-t}{\rVert}/^{2}}{ 2 l^2}) + \delta_{tt'}\sigma^2$. The hyperparameter $l$ of the squared exponential kernel defines a characteristic length scale above which data points are not correlated. The white noise is incorporated through a variance term $\sigma$ (an hyperparameter) with $\delta_{tt'}$ being the Kronecker function. 
The white noise should describe the pipeline error bars and the length scale should be of the order of one HST visit duration. Having a shorter length scale can allow us to reproduce shorter features providing a better correction of the systematics. However, it may lead to an over fitting.  GP are implemented using George \citep{Ambikasaran:2014aa}.

We note that Visit 0 only has one orbit, which makes the correction of the breathing effect ambiguous as its reproducibility between orbits of the same visit cannot be assessed. Therefore, we did not correct the fluxes from this visit.

\clearpage

\section{Other lines}
\begin{figure}
\centering
\includegraphics[width=\columnwidth]{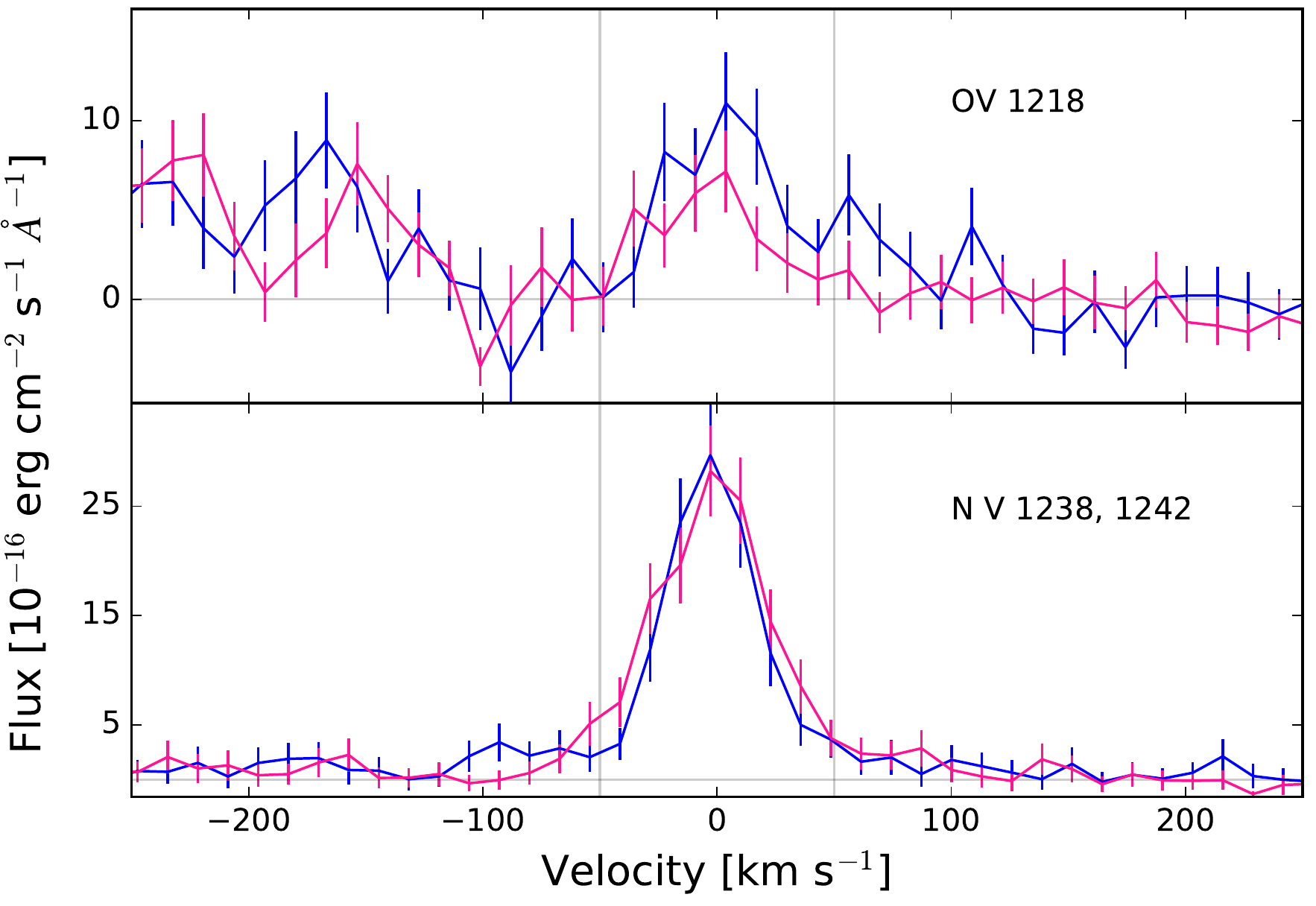}
\caption[]{\label{fig_others}Averaged out-of-transit (blue) and in-transit (pink) spectra of the \ion{O}{v} at $1\,218.344$~\AA {} (top panel) and the \ion{N}{v} doublet at $1\,238.8$~\AA  {} and $1\,242.8$~\AA {} (bottom panel) with both lines summed in velocity space. Vertical grey line indicates the $[-50,+50]$~\kms area.}
\end{figure}
\begin{figure}
\centering
\includegraphics[width=\columnwidth]{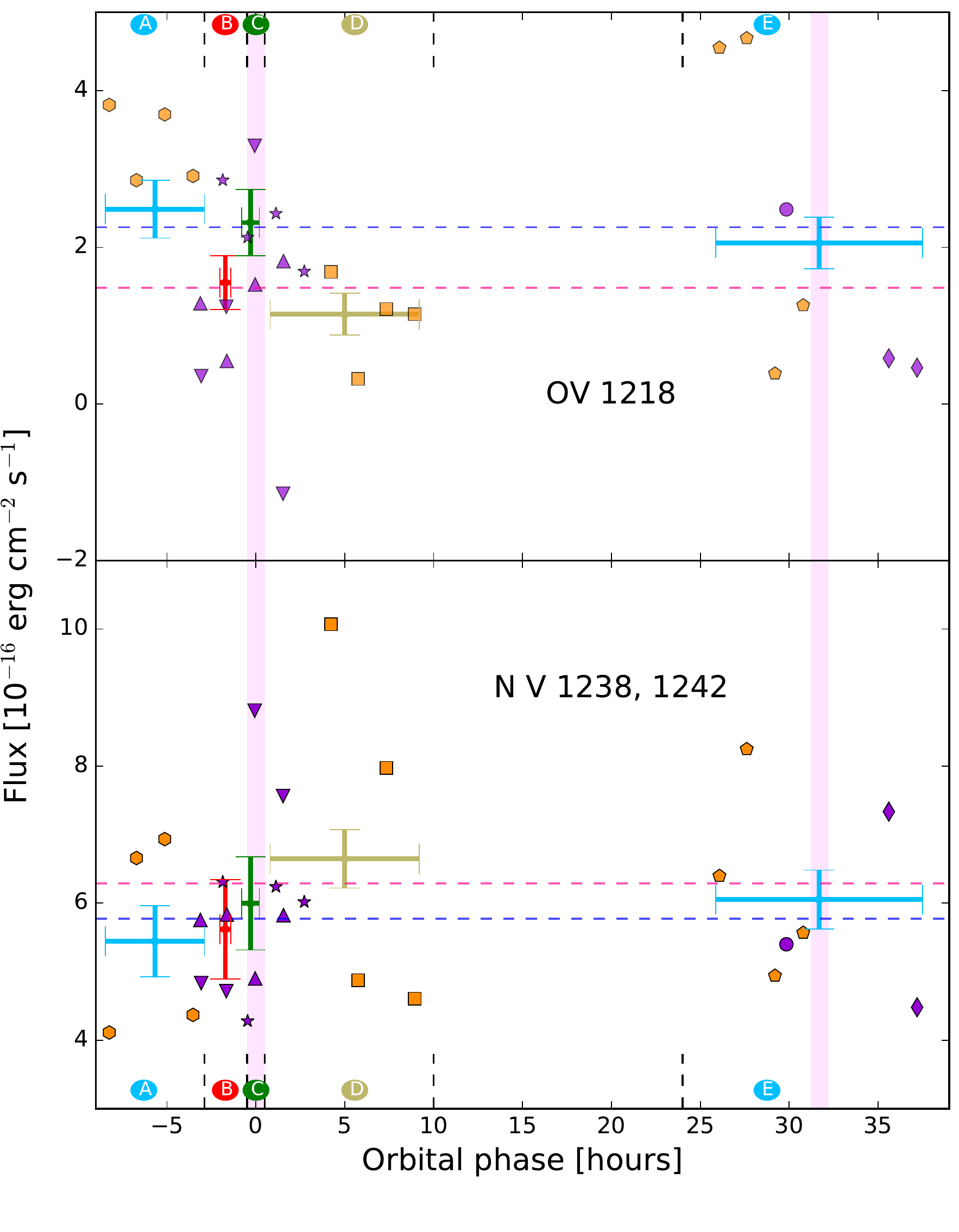}
\caption[]{\label{lc_others} Light curves of GJ~436  \ion{O}{v} line (top panel) and \ion{N}{v} (bottom panel) . Legend is the same as in Fig~\ref{lightcurve}. Fluxes are integrated for the different temporal regions (see Fig~\ref{lightcurve}). Horizontal dashed line indicates the out-of-transit flux (blue - regions A and E) and the in-transit flux (pink - regions B, C, and D). Symbols indicate the orbit fluxes (see Table~\ref{tab:data})}
\end{figure}

\begin{table}
\begin{center}
\caption{\label{tab:otherlines} List of identified stellar lines in STIS/G140M.}
\begin{tabularx}{\linewidth}{ZZZZ}
\hline
\hline
\multirow{2}{*}{Species} & Wavelength & Stellar flux & Absorption \\
& [\AA] & ($10^{-16}$ erg cm$^{-2}$ s$^{-1}$) & \% \\
\hline
\ion{Si}{iii} & 1206.510 & $3.8\pm0.3$ & $47\pm 10$ \\
\ion{O}{v} & 1218.344 & $2.2\pm0.2$ & $36\pm15$ \\
\multirow{2}{*}{\ion{N}{V}} & 1238.8 & \multirow{2}{*}{$5.5\pm0.3$} & \multirow{2}{*}{$-0.07\pm12$} \\
& 1242.8& &  \\
\hline
\end{tabularx}\\
\end{center}
\vspace{0.2in}
\end{table}

Apart from the L$\alpha$ line, we have identified several other lines : \ion{Si}{iii} at 1206.51 \AA, \ion{N}{V} at 1238.821, 1242.804 \AA,  and \ion{O}{V} at 1218.344 \AA. Those lines have a flux two magnitudes lower than the L$\alpha$ line. The flux measured during one single orbit is in a photon-starved regime and can present strong variations of magnitude (even a negative flux) and shape from one orbit to another. 
We therefore averaged  all the spectra within the orbital phase regions labelled A to E on Fig.~\ref{lightcurve} (cf.~Table~\ref{tab:defregion}): (A) before the exospheric transit signature in the blue wing of \lya, (B) during the exospheric transit ingress, (C) during the optical transit, (D) during the exospheric transit egress, and (E) after the exospheric transit. Regions (B), (C), and (D) are considered `in-transit'; regions (A) and (E) are `out of transit'. We consider the flux integrated in the velocity band [-50,50]km s$^{-1}$. Let $F_{in}$ and $F_{out}$ be those fluxes during the transit and out of the transit respectively (dashed lines in Fig~\ref{lc_others}). The absorption is defined as $1-F_{in}/F_{out}$. Table \ref{tab:otherlines}.1 shows results for each line. The \ion{N}{V} lines are summed together in the velocity space. No absorption is observed for this specie. 
An absorption signal is measured in the \ion{O}{V} line, but it may not be related to the planet as only the ingress and egress fluxes are absorbed. Finally, an absorption signal in  \ion{Si}{iii} line correlated with the planet transit is detected at almost five sigma (see Sect.~\ref{silicon_result}).

\end{appendix}


\end{document}